\documentclass{article}
\usepackage[utf8]{inputenc}
\usepackage{graphicx}
\usepackage{authblk}
\usepackage{siunitx}
\usepackage{xcolor}
\usepackage{hyperref}

\begin{document}

\title{Engineering the sub-Doppler force in molecular magneto-optical traps}
		
\author[1$\star$]{S. Xu}
\author{P. Kaebert}
\author{M. Stepanova}
\author{T. Poll}
\author[1$\star$]{M. Siercke}
\author{S. Ospelkaus}	

\affil{Institut f\"ur Quantenoptik, Leibniz Universit\"at Hannover, 30167~Hannover, Germany}
\date{March 2022}
\affil[$\star$]{s.xu@iqo.uni-hannover.de} 
\affil[$\star$]{siercke@iqo.uni-hannover.de}

\maketitle

Magneto-optical trapping has been at the forefront of atomic and molecular cooling experiments for over three decades \cite{cite12,cite13}. The technique is used in virtually every lab working with ultracold neutral particles, owing to the fact that it is able to capture, trap and cool species down to temperatures inaccessible by any other, non-optical method. Once captured and cooled in a MOT, other techniques such as sub-Doppler cooling in molasses \cite{cite18, cite19, cite10, cite05, cite23, cite24} and evaporative cooling \cite{cite20, cite21, cite22, cite11} can be used to reduce temperatures and increase phase space densities even further. Sub-Doppler mechanisms are, in fact, present even during the MOT stage. Their effect on MOT temperatures can be beneficial \cite{cite01} or negligible \cite{cite14} or, as is the case in current molecular MOTs, sub-Doppler effects can significantly increase MOT temperatures \cite{cite06,cite07,cite08,cite09}. As such, the high temperatures are significantly limiting the phase space density of molecular systems. In this paper we propose experimental techniques to engineer these sub-Doppler effects. In particular we illustrate this by theoretically engineering a molecular magneto-optical trap for CaF, where the sub-Doppler heating has been turned into sub-Doppler cooling. Monte Carlo simulations of the molecular cloud show a significant decrease in temperature and a consequential increase in density, which will have considerable impact on experiments using laser coolable molecules attempting to reach quantum degeneracy.

In a magneto-optical trap, particles are subjected to light from six laser beams, incident from six directions, as well as a quadrupole magnetic field. The purpose of the laser beams is two-fold: $1)$ their frequencies are tuned in such a way as to produce a velocity damping force on the particles, and $2)$ together with the magnetic field they produce a spatial restoring force. In the case of a type-I MOT operating on a $J \rightarrow J+1$ transition \cite{cite12}, as is used to trap most atomic species, simply red detuning the lasers from resonance and choosing the correct circular polarizations ensures both damping and trapping. In the case of a molecular, type-II MOT \cite{cite26} operating on a $J \rightarrow J-1$ transition, multiple laser frequencies and polarizations must be used \cite{cite06,cite07,cite08,cite09}. In both cases the resulting photon scattering produces cooling and trapping. This stands in stark contrast to the forces responsible for sub-Doppler cooling/heating. Here, photon scattering serves the role of optical pumping, and the scattering itself is not the main contributor to the momentum change \cite{cite19}. Rather, sub-Doppler processes generally rely on differential AC Stark shifts for slowing. While the probability to scatter a photon scales as $I/\delta^2$, where $I$ is the laser intensity and $\delta$ the laser detuning from resonance, the AC Stark shift scales as $I/\delta$. This, then, highlights the idea behind this paper: at least to some degree, experiments should be able to control Doppler forces, relying on scattering, and sub-Doppler forces relying on the AC Stark shift, independently by adjusting laser intensity and detuning or adding additional lasers.

\begin{figure}%
    \begin{center}
     \includegraphics[scale=0.47]{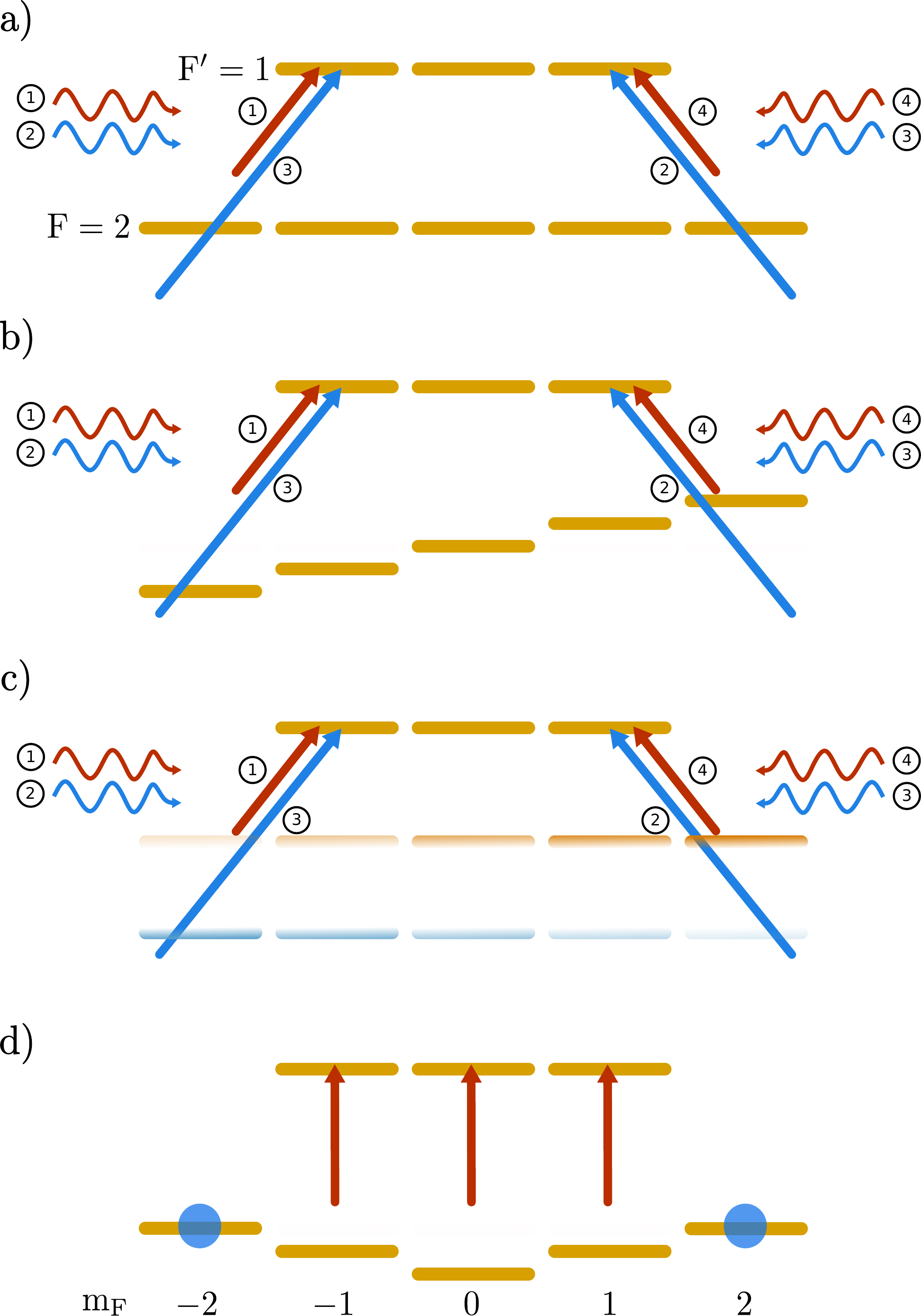}
     \caption{1D MOT laser configuration for a $F=2$ to $F'=1$ system. \textbf{a)}Laser frequencies, directions and polarizations needed. The four laser beams are labeled 1 through 4 respectively with their color representing their detuning and their horizontal direction indicating their polarization. \textbf{b)} In a positive magnetic field the ground state Zeeman shift causes light to preferentially be scattered from beams 3 and 4, resulting in a momentum kick to the left. \textbf{c)} Transitions as seen by the lasers for a molecule with a velocity to the right. Lasers 3 and 4 will see red shifted transitions (red ground state levels) while 1 and 2 see blue shifted ones (blue ground state levels). The overall laser detuning is chosen such that laser 4 comes into resonance sooner than laser 2, resulting in a damping force to the left. \textbf{d)} AC Stark shift and optical pumping in red-detuned cork-screw molasses. }
    \label{fig:1}
    \end{center}
\end{figure}

To illustrate this point we focus for the rest of the paper on the dual-frequency type-II MOTs currently used in ultracold molecule experiments. In particular, we focus on the case of CaF molecules. The ground state of CaF has a $F=2$ level which couples to the $F'=1$ level in the excited state (figure \ref{fig:1}). Furthermore, the excited state g-factor is small compared to that of the ground state. To produce a magnetic restoring force in such a system, two laser frequencies are necessary (figure \ref{fig:1}a): a red-detuned and a blue-detuned frequency as was first pointed out in \cite{cite02}. For the beams travelling to the left (beams 3 and 4), the red-detuned beam is $\sigma^-$ polarized and the blue-detuned beam $\sigma^+$. When the molecules move to a region of higher field (to the right), the $m_F=\pm 2$ and $m_F=\pm 1$ levels shift closer to resonance with these beams (figure \ref{fig:1}b). Conversely, the beams propagating to the right are shifted out of resonance, and the resulting photon scattering imbalance pushes molecules towards the center of the quadrupole field.

If the detuning of the red and blue-detuned laser components is equal but opposite, there is no velocity dependent force: when a molecule travels towards one of the laser beam pairs it scatters more photons from the counterpropagating, red-detuned beam, but it also scatters more photons from the blue-detuned, co-propagating beam. To get a velocity-damping force in this system then, both frequency components need to be shifted slightly higher, so that the red-detuned component is closer to resonance than the blue-detuned one. This ensures that molecules preferentially scatter photons from the counterpropagating laser beams (figure \ref{fig:1}c). 

\begin{figure}%
    \begin{center}
     \includegraphics[scale=0.47]{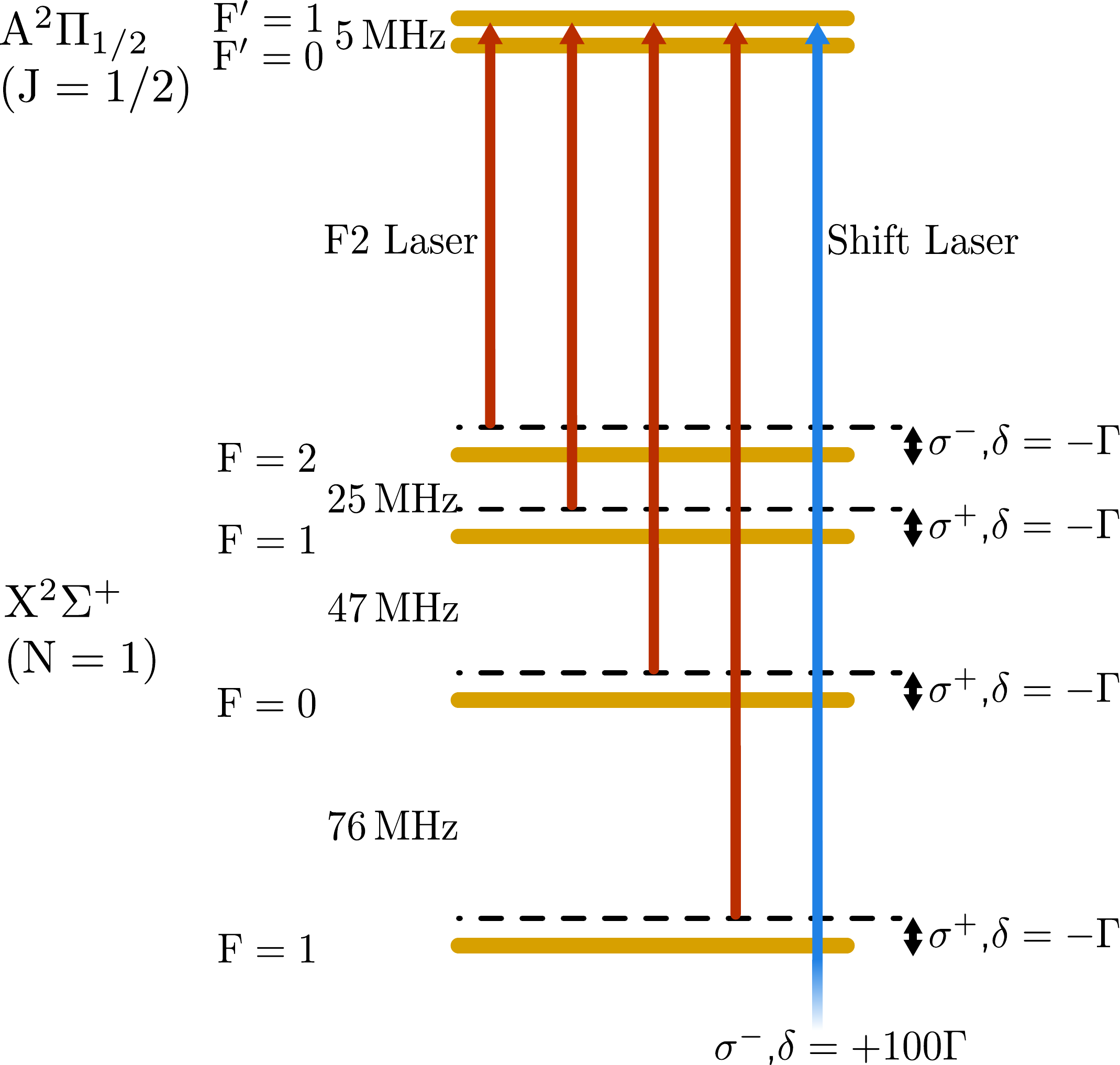}
     \caption{Ground and excited state hyperfine levels of CaF and laser frequencies used to address them. The ground states are coupled to the excited states with four laser frequencies, each detuned by $\delta=-\Gamma$ from their respective transition. Also shown in blue is the shift laser, detuned by $\delta=+100\Gamma$ and with polarization equal to that of the F2 laser.}
    \label{fig:2}
    \end{center}
\end{figure}

The laser configuration producing the dual-frequency MOT force brings with it an unfortunate consequence: near zero velocity, at the trap center, the red-detuned laser beams are closer to resonance than the blue-detuned ones. This results in AC Stark shifts that lower the energy of the ground states by an amount depending on the coupling strength. In 1D, the counterpropagating red-detuned $\sigma^+$ and $\sigma^-$ laser beams form a corkscrew polarization pattern, where the polarization is linear everywhere and rotates as a function of distance from the center \cite{cite03}. Choosing the quantization axis to be along this polarization, molecules near zero velocity will be preferentially pumped into the $F=2, m_F= \pm 2$ states, which have the least AC Stark shift and, due to the red detuning, the highest energy (figure \ref{fig:1}d). As a consequence, when molecules move at low velocities in this laser configuration, they will constantly "roll down" a potential hill as they move through the corkscrew polarization, causing sub-Doppler heating \cite{cite15}. 
In reality the level structure of CaF is more complicated (figure \ref{fig:2}). The $X^2\Sigma^+$ ground state consists of four hyperfine energy levels, $F=1$,$0$,$1$ and $2$, spaced by \SI{76}{MHz}, \SI{47}{MHz} and \SI{25}{MHz}. The $A^{2}\Pi_{1/2}$ excited state has a hyperfine splitting of \SI{5}{MHz}. Four laser frequencies are used to drive transitions from the four ground states to the excited states, each detuned by $\delta=-\Gamma$ with respect to the $F'=1$ excited state \cite{cite02}. The lasers driving the $F=1$ and $F=0$ ground state levels are $\sigma^+$ polarized while the one driving the $F=2$ ground state is $\sigma^-$ polarized. As such, the $F=2$ to $F'=1$ transition closely resembles the situation depicted in figure \ref{fig:1}.

\begin{figure}%
    \begin{center}
     \includegraphics{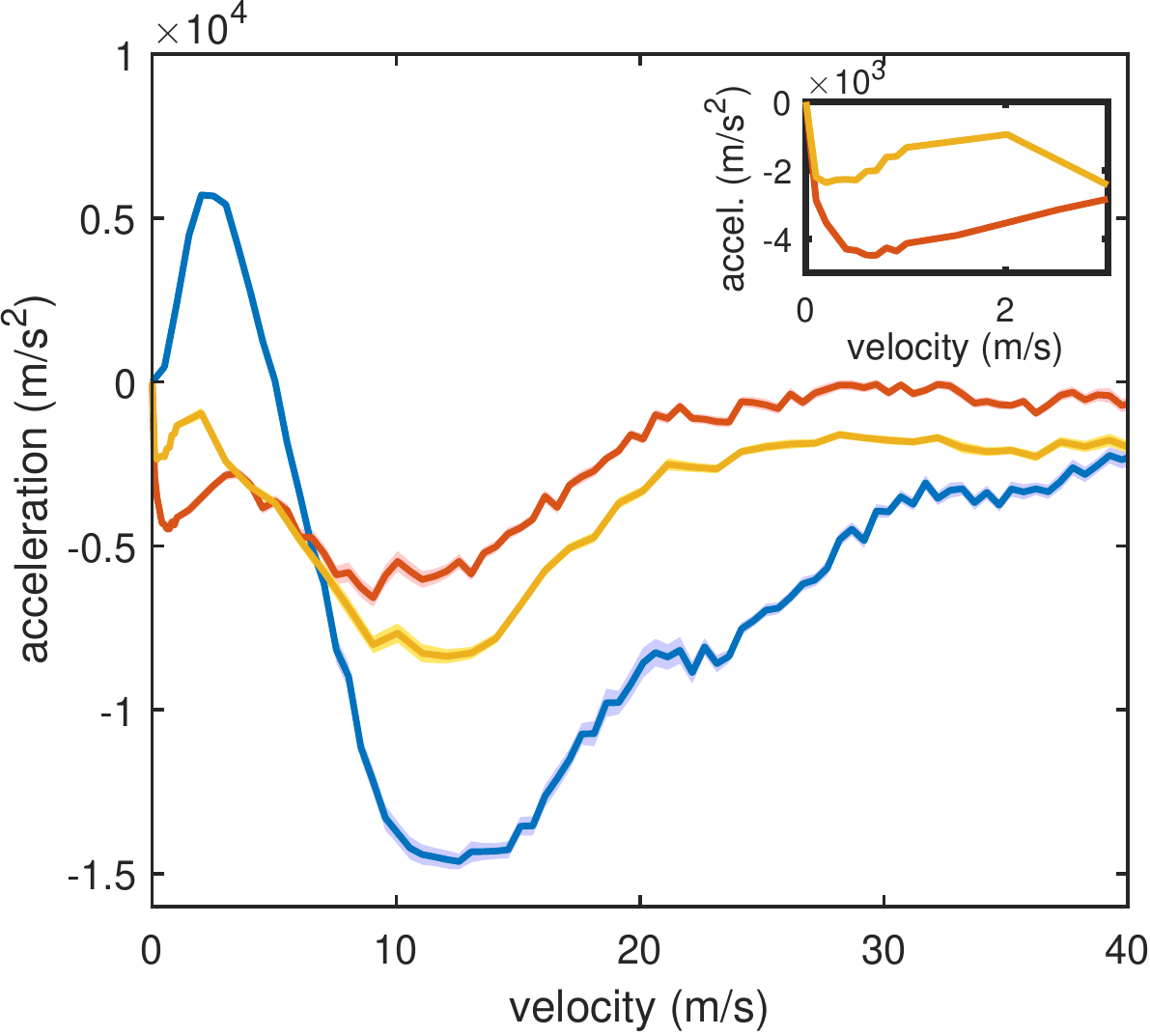}
     \caption{Acceleration vs velocity for molecules in the MOT. The blue curve shows the acceleration in the usual 4 laser component, dual-frequency MOTs used to date \cite{cite04}. The red curve shows the acceleration when the blue-detuned shift laser is added with polarization equal to the $F=2$ laser component, showing sub-Doppler cooling below \SI{3}{m/s}. The yellow curve depicts the acceleration when the blue laser polarization is reversed with respect to the $F=2$ component. Inset: closeup of the sub-Doppler cooling part of the red and yellow curves.}
    \label{fig:3}
    \end{center}
\end{figure}

A full 3D simulation of the 16-level optical Bloch equations (OBEs) for CaF, using this laser configuration confirms our intuitive picture of optical pumping into the $m_F = \pm 2$ states. The resulting velocity dependent force is depicted as the blue curve in figure \ref{fig:3}. The figure clearly shows the Doppler cooling and sub-Doppler heating features observed in experiment \cite{cite23}. To get rid of this sub-Doppler heating, and perhaps even turn it into cooling, the $m_F=\pm 2$ states must be shifted down in energy instead of up, while not affecting the photon scattering much. This can be accomplished most easily with a laser detuned far into the blue. Because $m_F$ is defined relative to the polarization of the most red-detuned frequency component (the one that is most responsible for optical pumping), this far off-resonant laser should have the same polarization as this component. For the rest of the discussion we shall refer to this blue-detuned laser as the shift laser and to the most red-detuned laser as the F2 laser (see figure \ref{fig:2}). For our simulations we choose a blue detuning of the shift laser of $100 \Gamma$ (angular frequency), where $\Gamma = 2 \pi \times \SI{8,3}{MHz}$ is the transition's natural linewidth. The red curve in figure \ref{fig:3} shows the result of the OBEs when including this shift laser with a power of $\SI{80}{mW}$ and a Gaussian beam diameter of \SI{4}{mm}. The other four, regular MOT frequency components share \SI{200}{\milli W} of power and have a beam diameter of \SI{20}{\milli m}. The sub-Doppler heating which previously dominated the force for velocities below \SI{6}{m/s} has disappeared and in its place there is a sub-Doppler cooling force below \SI{3}{m/s}. Applying the shift laser to the MOT furthermore seems to have reduced the Doppler force, presumably because the additional AC Stark shift has changed at what velocity the various ground states become resonant with the lasers, however the force is still more than strong enough to keep the molecular cloud cold.

\begin{figure}%
    \begin{center}
     \includegraphics{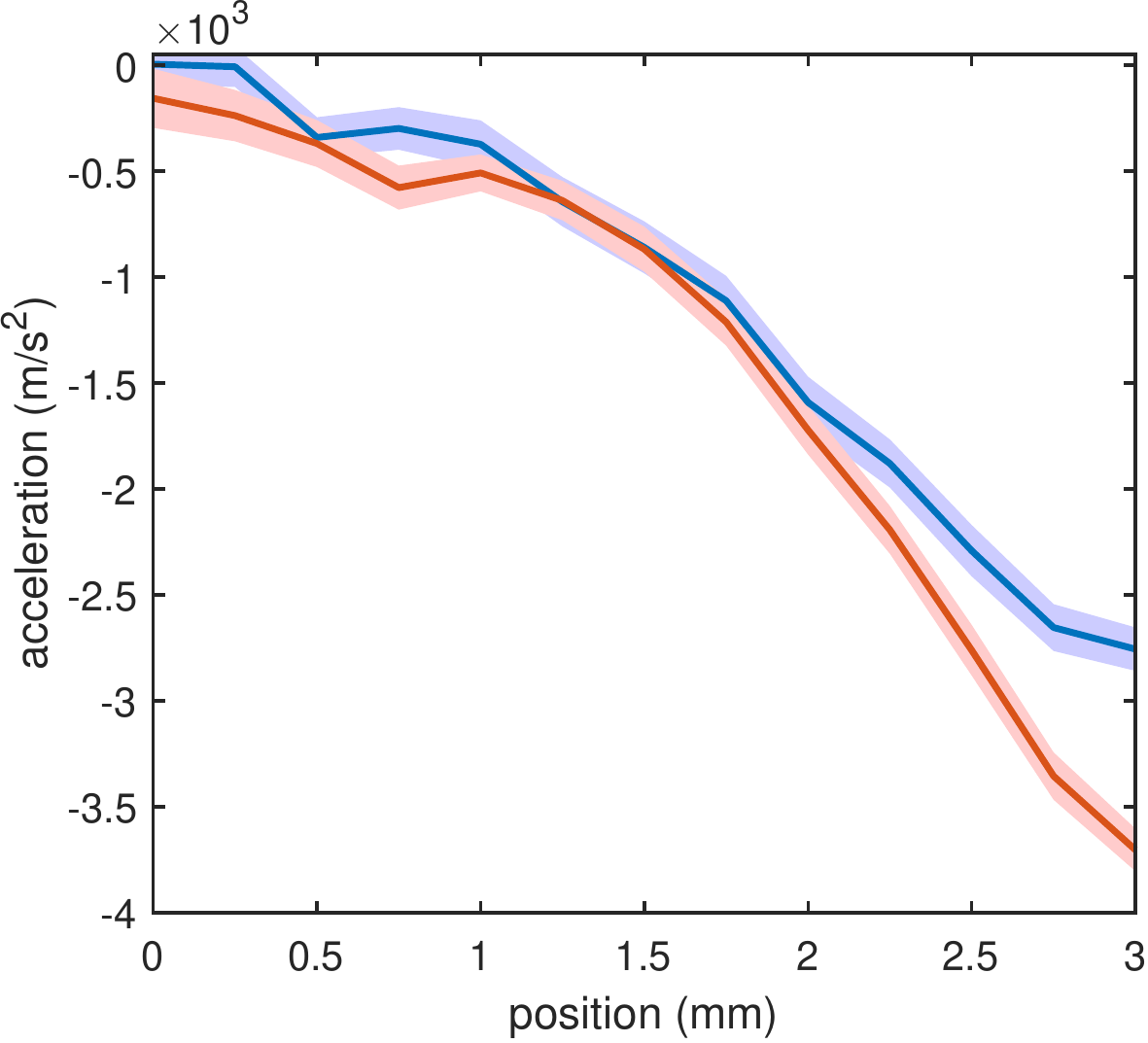}
     \caption{Acceleration vs position for molecules in the MOT. The blue curve shows the acceleration in the usual four laser component, dual-frequency MOTs used to date \cite{cite04}. The red curve shows the acceleration when the blue-detuned shift laser is added with polarization equal to the $F=2$ laser component. Adding the extra blue laser seems to have little effect on the trapping force. Shaded areas indicate the spread of values for different runs of the OBEs}
    \label{fig:4}
    \end{center}
\end{figure}

The effect of the shift laser on the trapping force is shown in figure \ref{fig:4}. A blue-detuned laser can in principle result in a repulsive force, but this effect is negligible for our parameters as can be seen from the figure. Since the radius of the shift laser is chosen to be \SI{2}{mm} the figure only shows the acceleration for positions up to \SI{3}{mm}. Calculating the trapping force using the OBEs needs a lot of averaging \cite{cite04} as it fluctuates a lot depending on the initial conditions chosen. As such, while the red curve in figure \ref{fig:4} seems to show a small force at the MOT center, our uncertainty in the calculation is consistent with zero force and we force it to be so in our further modelling.

The requirement that the shift laser has the same polarization as the F2 laser translates to the condition that the phase of the $\sigma^+$ and $\sigma^-$ parts should be the same over the size of the MOT. This results in two constraints for experiments (see supplementary material): the red and blue components must be co-propagating from the point on where their $\sigma^+$ and $\sigma^-$ beams are split, and the path length difference between counterpropagating beams must be an integer multiple of the beat length between the F2 and shift lasers. The former constraint is necessary for the polarization gradients of the two components to remain in-phase with each other. The latter constraint ensures that the polarization direction of the two components is the same over the size of the MOT. For the detuning chosen in our simulations, this constrains the path length difference between counterpropagating beams to multiples of \SI{36}{cm} with a tolerance of \SI{1}{cm} to keep the angle between F2 and shift laser polarization below $5^\circ$. Both constraints should be readily realizable in experiments. The yellow curve in (figure \ref{fig:3}) shows what happens to the acceleration when the polarization of the shift laser is reversed (i.e. $\sigma^+$ turned to $\sigma^-$ and vice versa). This causes the polarizations of the shift and F2 lasers to not be equal at every point in space anymore, significantly reducing the sub-Doppler cooling. There does remain a small cooling feature however, and as such, the shift laser should help improve MOT temperatures and densities even when the above two constraints are not met.

Using our results from the OBEs we can estimate the temperatures and densities the sub-Doppler cooling MOT should be capable of reaching. To this end, we perform a simple Monte Carlo simulation (see supplementary materials). From the velocity distribution obtained this way  (see figure \ref{fig:sup2}) we estimate the MOT temperature to be \SI{40}{\micro K}, well below the Doppler limit of $\approx\SI{200}{\micro K}$ \cite{cite05}. Similarly low MOT temperatures have previously been observed in experiment for blue-detuned MOTs of atoms \cite{cite01,cite16}, while current dual-frequency CaF MOTs only reach temperatures of about \SI{1}{\milli K}, and, consequently, densities on the order of \SI{1e6}{cm^{-3}}\cite{cite17}. The peak density of our MOT is simulated to be \SI{4e8}{cm^{-3}} for \SI{1e5}{} molecules, a value much higher than current molecular MOTs are capable of reaching. It should be noted that the density is limited in part by our choice of magnetic field gradient (\SI{15}{G/cm})  and higher densities may be reached by increasing this value.

In conclusion, we have shown that it is possible to engineer sub-Doppler forces inside a Magneto-Optical trap using far off-resonant laser beams and a careful choice of polarization. Using this principle we designed a MOT for CaF molecules with sub-Doppler cooling and estimated that it should reach temperatures of \SI{40}{\micro K} and densities of \SI{4e8}{cm^{-3}} for $10^5$ molecules. Such densities are much larger than is currently being reached in molecular CaF MOTs \cite{cite09}, and they are a direct consequence of the lower MOT temperature. The low temperature and high density result in a phase space density of our simulated MOT of \SI{2e-8}{}. While this paper has concentrated on improving molecular magneto-optical traps, the idea that AC Stark shifts, and as such, sub-Doppler effects, can be engineered is quite general, and we expect many uses for the technique across the broad spectrum of atom and molecular physics experiments being performed today. 

\section*{Acknowledgements}

 We gratefully acknowledge financial support through  Germany’s Excellence Strategy – EXC-2123/1 QuantumFrontiers.

\bibliography{main}
\bibliographystyle{iopart-num}

\newpage
\section*{Supplementary Materials}
\subsection*{Monte Carlo Simulation}

\begin{figure}%
    \begin{center}
     \includegraphics{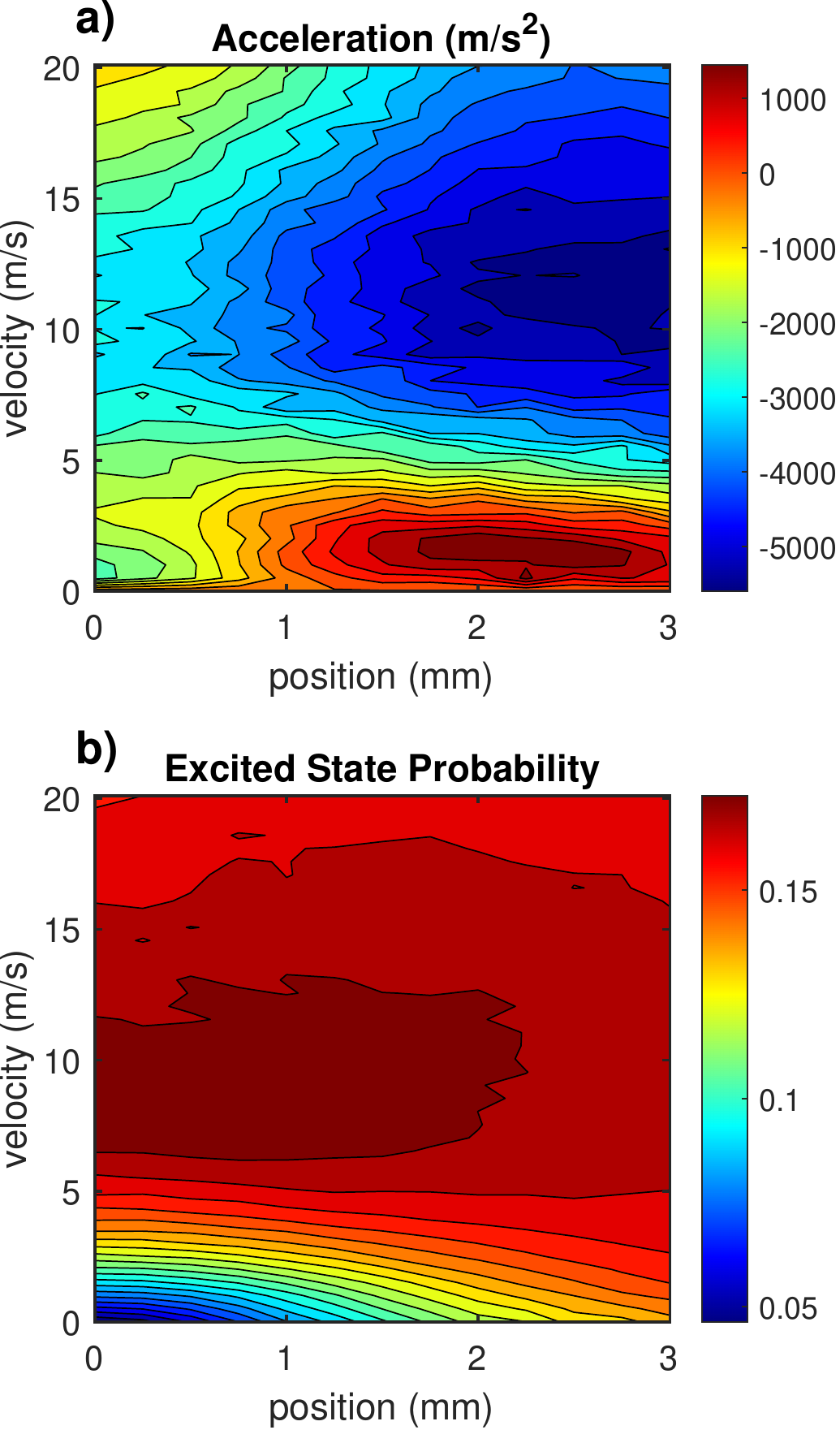}
     \caption{Velocity damping force (acceleration) (a) and excited state probability (b) of the molecules vs position and velocity in the MOT.}
    \label{fig:sup1}
    \end{center}
\end{figure}

Here we present the details of our Monte-Carlo simulation. The optical Bloch equations are solved to find the damping and trapping forces, closely following the method outlined in \cite{cite04}. To account for the effects of vibrational repumping, we reduce all forces by a factor of 2. For simplicity we use the axial 1D trapping force with the shift laser present and ignore the fact that the MOT is weaker in the radial direction. We calculate the velocity damping force as well as the probability of being in the excited state for velocities up to $4\Gamma/k$ and positions up to \SI{3}{mm} (figure \ref{fig:sup1}). Here, $\Gamma$ is the natural linewidth of the transition and $k$ the magnitude of the wave vector. Since the shift laser beam diameter is smaller than the other MOT lasers we see a transition from sub-Doppler cooling to sub-Doppler heating at larger positions. Even though our initial cloud size overlaps significantly with the heating region, the final MOT size is smaller than the shift laser beam, so that the heating does not affect the final cloud temperature. For the trapping force we make the approximation that its dependence on the velocity is negligible, and use the force in figure \ref{fig:4}. As is stated in the main text, the force in figure \ref{fig:4} has a small value at the MOT center, which would be unphysical. Our error estimates show that this value is consistent with zero, however, and we set it to zero for our simulations.

\begin{figure}%
    \begin{center}
     \includegraphics{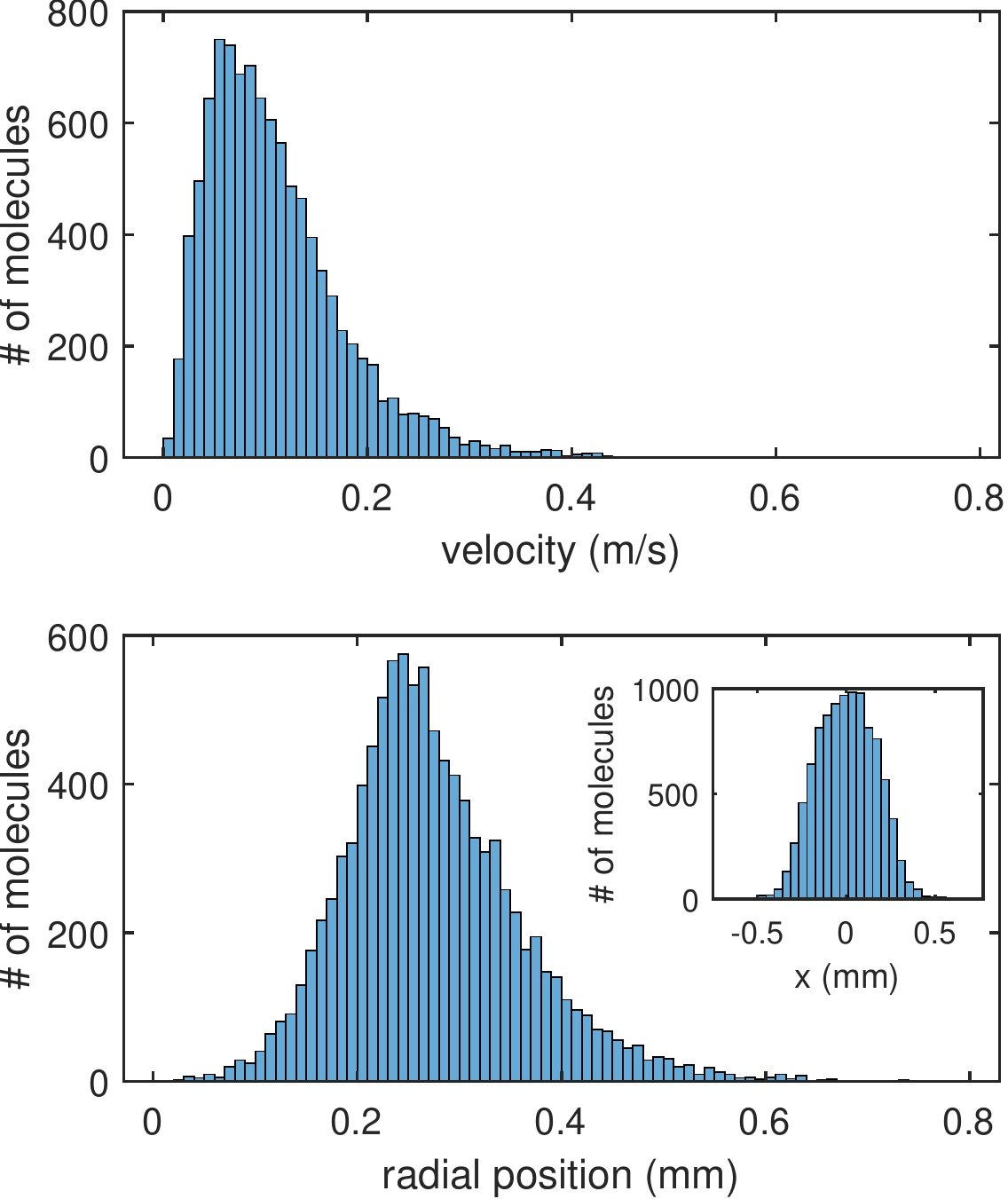}
     \caption{Distribution of velocities and radial positions of $10^4$ molecules subjected to the forces given in figures \ref{fig:4} and \ref{fig:sup1}. The inset shows the 1D distribution of molecules along one of the Cartesian coordinates.}
    \label{fig:sup2}
    \end{center}
\end{figure}

We start the simulation with a uniform 3D distribution of $10^4$ molecules with radius \SI{3}{mm}. The 3D velocity distribution is also chosen to be uniform, with a radius of \SI{1}{m/s}. Because the profiles in figures \ref{fig:4}
and \ref{fig:sup1} are relatively coarse we use linear interpolation of their data to calculate the force at the velocity and position of each molecule. Photon scattering is included as momentum kicks of $\hbar k$ in a random direction with a probability of $\Gamma \rho_{ee} dt$, where $\rho_{ee}$ is the excited state probability and $dt$ is the timestep of the calculation. In case this probability exceeds $1$ we instead give the molecule a momentum kick of $\sqrt{\Gamma \rho_{ee} dt}$ every timestep. Time steps in the simulation are chosen to switch periodically between $100/\Gamma$ and $1/\Gamma$. The $100/\Gamma$ steps allow for the molecules to propagate in space while keeping the computation time reasonable, while the $1/\Gamma$ steps more accurately model photon scattering. For the final time evolution we give $10^6$ steps of $1/\Gamma$ to make sure equilibrium is reached. The resulting velocity and radial position distributions are plotted in figure \ref{fig:sup2}. From these we extract a temperature of \SI{40}{\micro K} using $T=m v_{rms}^2 /3\,k_{\textrm{B}}$. The central density is calculated by simply counting the number of molecules in a $0.3 \times 0.3 \times 0.3$\,\SI{}{mm^3} cube in the center of the cloud.

\subsection*{Requirements on red and blue laser components}
For the conventional MOT, at low velocities, molecules are optically pumped by the red most laser component (F2 laser), which forms a corkscrew polarization pattern in space. Choosing the quantization axis along the local polarization of this laser, the OBEs confirm that molecules are pumped into the $m_F=\pm2$ ground state levels, which are dark states to this laser. In order for our scheme to produce sub-Doppler cooling, our blue AC Stark shift laser must shift the $m_F=\pm1$ and $m_F=0$ levels above the $m_F=\pm2$ levels, and as such it needs the same local linear polarization as the F2 laser. Two conditions can be derived from this requirement. First, while it is okay that the local polarization changes in time (due to e.g. vibrations of the mirrors), the corkscrews produced by the blue and red laser must not dephase with respect to one another. The easiest way to achieve this in experiment is to have both lasers overlap from the moment that the $\sigma^+$ and $\sigma^-$ components are split. This way, any phase induced by vibrations will be placed on both beams and their polarizations will rotate together.
Even with the blue and red lasers taking the same path, there is a second requirement: if the arm lengths for the $\sigma^+$ and $\sigma^-$ beams are the same there will be zero phase difference between them when they meet in the MOT center. If the arm length is different, however, the phase between $\sigma^+$ and $\sigma^-$ will be given by 
\begin{equation}
    \Delta \theta = k (z_+-z_-) = k \Delta z
\end{equation}
where $k$ is the wave vector and $z_+$ and $z_-$ are the path lengths for the $\sigma^+$ and $\sigma^-$ components respectively. Because shift and F2 lasers have different wave vectors, this phase is not necessarily the same for them, causing their linear polarizations to be at an angle with respect to each other. This phase difference, then, is given by

\begin{equation}
    \Delta \phi=\Delta k \Delta z = \frac{2 \pi (f_b - f_r)}{c} \Delta z = \frac{100 \Gamma}{c} \Delta z,
\end{equation}
using the detuning we have chosen for the shift laser in our simulations. From this, using the excited state linewidth of CaF, we see that the path length difference between counterpropagating MOT beams should be multiples of \SI{36}{cm}. If we allow for a $5^\circ$ angle between the polarizations of the red and blue beam we may have a path length difference uncertainty of $\pm$\SI{1}{cm}. 

Making sure that each counterpropagating MOT pair travels the same distance $\pm$\SI{1}{cm} should certainly be possible in experiments. If, on the other hand, one wants to use a single laser to produce all six MOT beams in order to not split power into multiple paths, care must be taken that each retro-reflected beam has traveled multiples of \SI{36}{cm}, a number that also seems reasonable.

\end{document}